\documentclass[final,5p,times,twocolumn]{elsarticle}

\usepackage{amssymb}
\usepackage{amsmath}

\journal{NIM A}

\begin{document}

\begin{frontmatter}



\title{Monte Carlo Simulations of the Transition Radiation Detector of the AMS-02 Experiment\tnoteref{label1}}
\tnotetext[label1]{\textcopyright{} 2017. This manuscript version is made available under the CC-BY-NC-ND 4.0 license, http://creativecommons.org/licenses/by-nc-nd/4.0/.}


\author{T.~R\"aih\"a}
\ead{tomi.raiha@physik.rwth-aachen.de}

\author{A.~Bachlechner}
\author{B.~Beischer}
\author{C.~H.~Chung}
\author{H.~Gast}
\author{S.~Schael}
\author{T.~Siedenburg}
\address{I. Physikalisches Institut B, RWTH Aachen University, 52056 Aachen, Germany}

\begin{abstract}

  The Transition Radiation Detector of the AMS-02 experiment on the
  International Space Station is used for the separation of cosmic-ray
  positrons and electrons from protons and anti-protons, and for the
  identification of nuclei up to carbon ($Z\le6$). We present the
  Geant4 simulation that is used to describe the ionization and
  transition radiation processes and compare its results to flight
  data from AMS-02. After applying empirical corrections to the
  simulated data, the particle energy deposition and likelihood
  distributions in the TRD are described with high accuracy.
\end{abstract}


\end{frontmatter}

\section{Introduction}
\label{sec:introduction}

The effect of Transition Radiation (TR) can be used to effectively separate light
and heavy particles in a high-energy physics detector. TR
photons are emitted when a relativistic charged particle crosses
boundaries between materials of different permittivity. Transition
Radiation Detectors (TRD) exploit the fact that the yield of TR
depends on the Lorentz factor $\gamma$ of a particle.

For the design of a TRD and advanced data analysis, the availability
of a realistic detector simulation is essential. Here, we compare data
recorded by the TRD \cite{2013NIMPA.706...43K} of the Alpha Magnetic
Spectrometer (AMS-02) \cite{amsdetector}, a general purpose particle
detector for cosmic rays mounted on the International Space Station
(ISS), to a Monte Carlo simulation. The AMS-02 TRD is capable of
separating protons from positrons with a rejection power of more than
10000 at a particle momentum of $10\,\mathrm{GeV}/c$ and 10 at
$1\,\mathrm{TeV}/c$~\cite{trdrejection}. With the current high voltage
settings, its dynamic range allows for the measurement of primary
ionization for ions up to carbon ($Z\leq6$). AMS-02 has collected a
vast amount of high-quality data since its launch in May 2011, which
can be used for optimization and verification of the detector
simulation.

The Monte Carlo simulation (MC) of the TRD is based on the widely used
Geant$4$ simulation toolkit \cite{geant4a_2003NIMPA.506.250A,
  geant4b_2006ITNS.53.270A}. We have used version 4.10.1. The
ionization of charged particles in the thin gas inside the TRD straws,
the production of TR in the fleece radiator and the photo-effect of TR
X-ray photons in the gas are managed by Geant4 classes as well. We
have optimized and tested the TRD simulation based on more than five
years of AMS-02 data for all relevant particle species that cover a
wide momentum range needed in data analysis. Despite the necessity for
a good description of the TR production and ionization energy losses
in thin absorbers, few studies have been published on experimental
validation of TRD simulation so far \cite{trdsim_2007NIMPA.583.485B,
  2000NIMPA.453..597A, 2002NIMPA.484..573G, 2006NIMPA.563..299G,
  2002PhLB..525..225G}. We describe the design of the AMS-02 TRD and
detail the steps to obtain an accurate simulation in
Section~\ref{sec:detector}. The most relevant AMS-02 TRD distributions
are compared to results from the Geant4 simulation in
Section~\ref{sec:comparison}, and conclusions are given in
Section~\ref{sec:conclusions}.

\section{Design and simulation of the AMS-02 TRD}
\label{sec:detector}

The TRD has approximate dimensions of $2.5 \times 2.5 \times
0.7$~m$^3$. It is located in the upper part of AMS-02 above the
permanent magnet that surrounds the inner tracker. The TRD provides
important information for the particle identification and allows the
selection of clean events through the recognition of secondary tracks
that can emerge from possible hadronic interactions occuring above and
inside the TRD.

\begin{figure}[!]
\begin{center}
\includegraphics[width=0.5\textwidth]{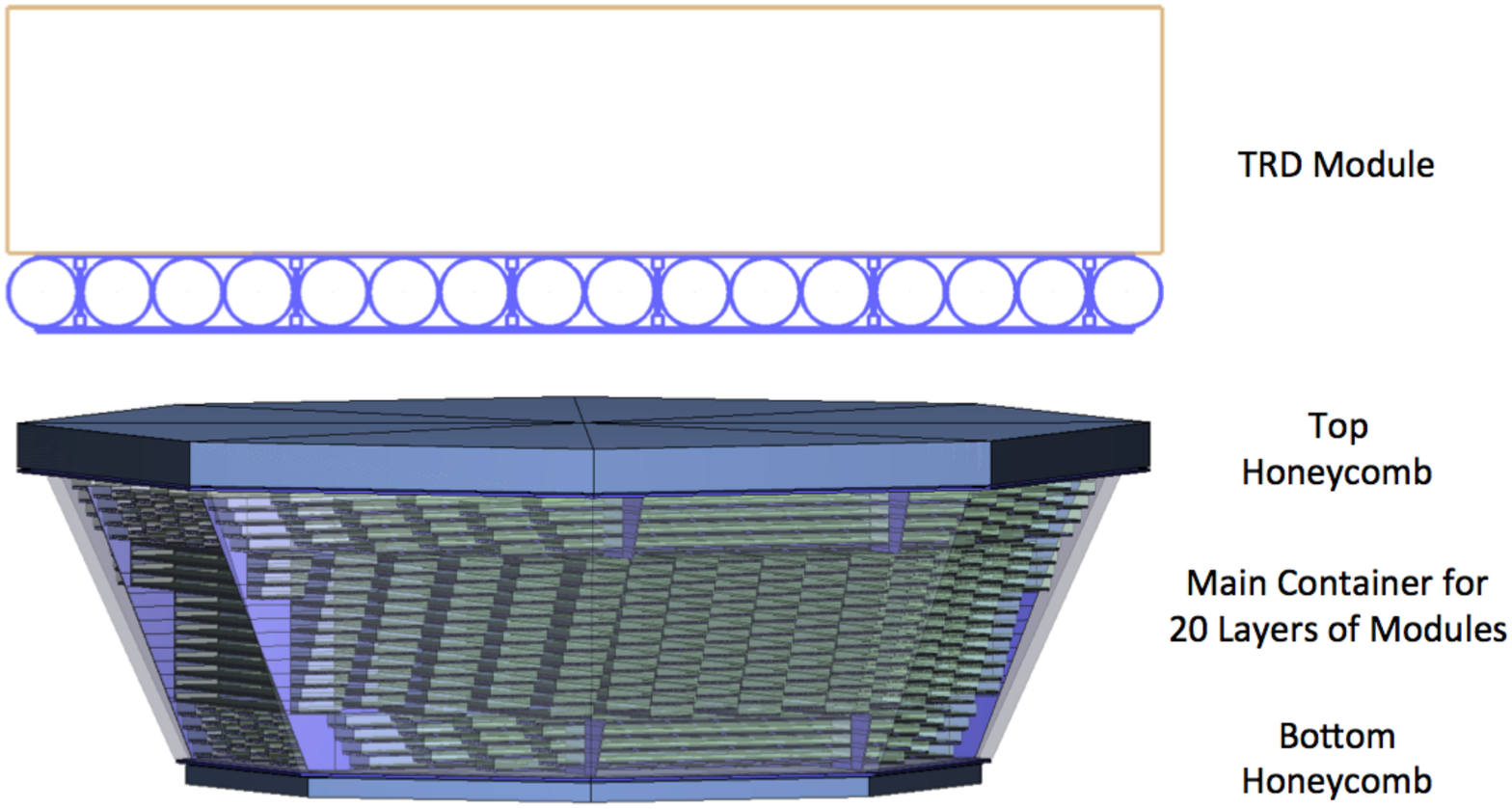}
\includegraphics[width=0.5\textwidth]{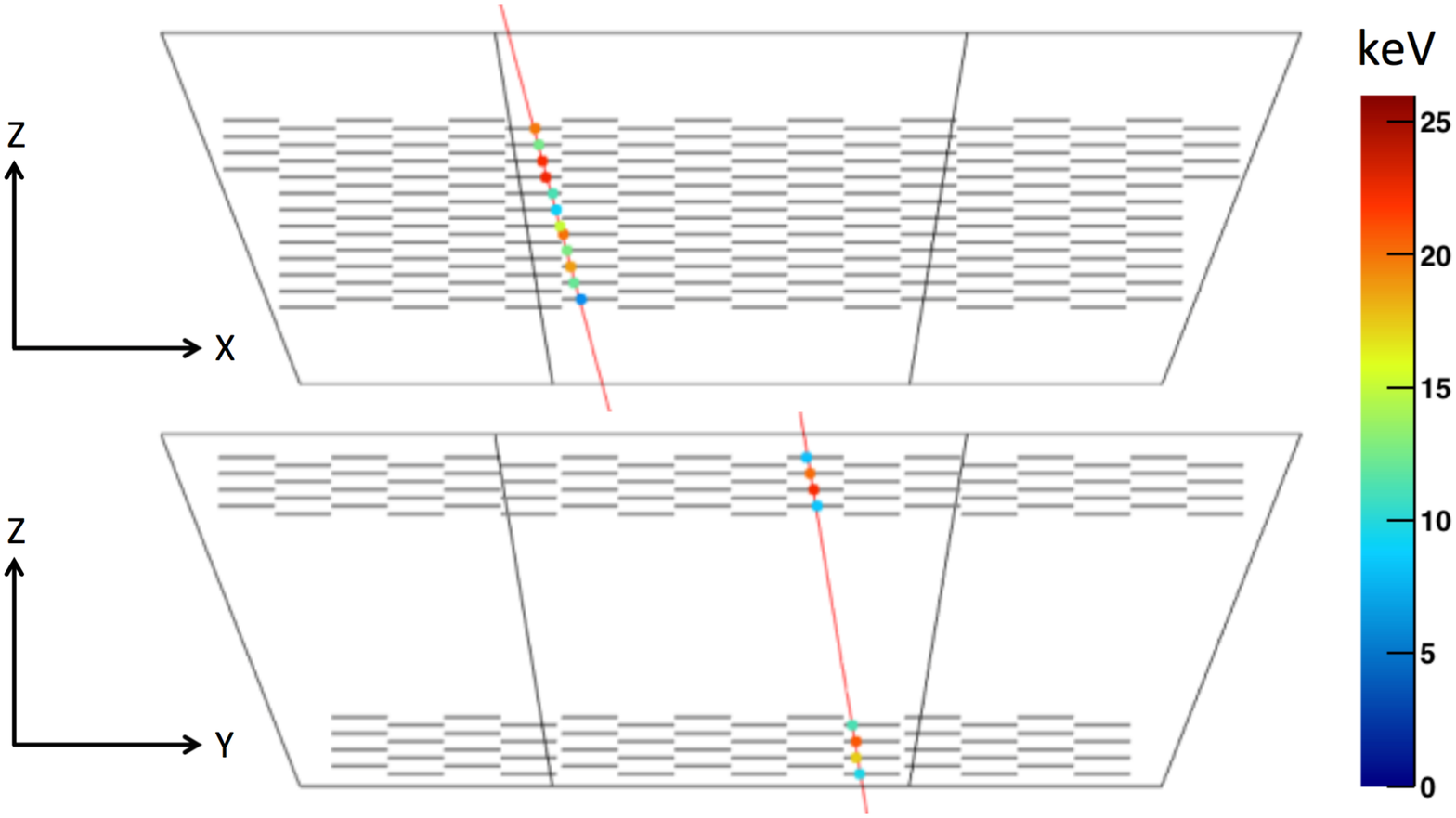}
\caption{Top: The model of the AMS-02 TRD. Bottom: A simulated
  $10\,\mathrm{GeV}/c$ electron event in the TRD viewed in the front
  (XZ) and side (YZ) projections.}
\label{fig:trd}
\end{center}
\end{figure}

The TRD consists of 20 layers of straw tubes with $0.6$~cm inner
diameter which are filled with a Xe/CO$_2$ ($91.5$\%/$8.5$\%) gas
mixture at a pressure of $1$~bar. The four topmost and four bottommost
layers are oriented perpendicularly with respect to the inner layers
to enable three-dimensional tracking with the TRD. The straws are
grouped into modules, each module containing $16$ straws and a 20\,mm
thick irregular fleece radiator. The radiator is made of polyethylene
and polypropylene fibers, with a density of $0.06$~g/cm$^3$. The
straws are made of double layer kapton-aluminum foil of $72 \, \mu$m
wall thickness, a gold-plated tungsten anode wire of $30 \, \mu$m
diameter attached in the center. In total, the TRD comprises $5248$
straws, with lengths varying between $1$--$2$~m. The model of the TRD
implemented in the simulation~\cite{schmanau_thesis} is visualized in
Fig.~\ref{fig:trd}.

The TRD is operated at a high voltage of 1.4\,kV for a gas gain of
$\sim3000$. The readout electronics is based on two VA-chip analog
multiplexers and 12-bit analog to digital converters (ADCs).  On
arrival of a trigger signal from the AMS main computer, the digitized
data are collected, reduced, and recorded~\cite{chung..electronics}.

We use Geant4 to simulate the passage of particles through the
TRD. For the simulation of ionization in the thin gas mixture inside
the TRD straws the Photo-Absorption Ionization (PAI) Model
\cite{1980ARNPS..30..253A} as implemented in Geant4 provides the most
accurate description of energy losses. The production of TR in the
fleece radiator is described best by the Geant4
\texttt{G4GammaXTRadiator} class~\cite{2004NIMPA.522..122G}, for which
the adjustable parameters were optimized already in
Ref.~\cite{trdsim_2007NIMPA.583.485B}. The optimal fiber thickness for
the production of TR was found to be $12$~$\mu$m, which is close to
its nominal value of $10$~$\mu$m. The relative fluctuation parameters
for the fiber thickness and the gap width between adjacent fibers were
set to $5$ and $0.66$, respectively. The thickness of the fleece
radiator blocks was adjusted to $21$~mm, resulting in a total weight of
$\sim\!60$~kg, in agreement with the measured weight. For the realistic
simulation of the absorption of TR X-ray photons in gas, the atom
de-excitation process was activated using the
\texttt{G4UAtomicDeexcitation} class. The production cuts for
secondary photons and electrons were 1\,mm outside the tubes and
0.001\,mm inside the tube walls. Apart from the processes used for
the description of TR and ionization in the TRD gas, we use the
standard electromagnetic physics list. For hadrons, we use the
standard elastic scattering processes, combined with the
\texttt{QGSP\_BIC} inelastic model. Details on the physics list used
for helium and heavier ions are given in Ref.~\cite{Aguilar2015a}.

The total energy deposited in a TRD straw is converted to a number of
ADC counts by the readout electronics chain. The ADC dynamic range
extends to $4095$~ADC counts, which corresponds to an energy deposition of
$85$~keV. An example of a $10\,\mathrm{GeV}/c$ electron simulated in
the TRD is illustrated in Fig.~\ref{fig:trd}. In the standard
analysis, a track reconstructed by the silicon tracker of AMS-02 is
extrapolated through the TRD to select hits on the track and to
determine the pathlengths inside the TRD straws. In addition, a
standalone TRD tracking is available that identifies tracks and
calculates pathlengths completely independent from the rest of
AMS-02. A particle passing through the complete TRD produces $18.5$
hits on average.

Before simulated data and data from the AMS-02 TRD can be compared,
the latter have to be calibrated to correct for time-dependent effects
from the constantly changing thermal environment on the
ISS~\cite{calib_heil,calib_gast}. In the first step, the relative
movement of the TRD with respect to the inner tracking system of
AMS-02 is reconstructed and corrected for. Then, a relative gas-gain
calibration factor is determined for every TRD module as a function of
time. This is done by selecting a clean sample of protons and fitting
the position of the peak of the Landau-like distribution of their
energy depositions. The same procedure is applied to simulated data,
to introduce realistic statistical fluctuations in the calibration
factors.

\begin{figure}[!]
\begin{center}
\includegraphics[width=0.5\textwidth]{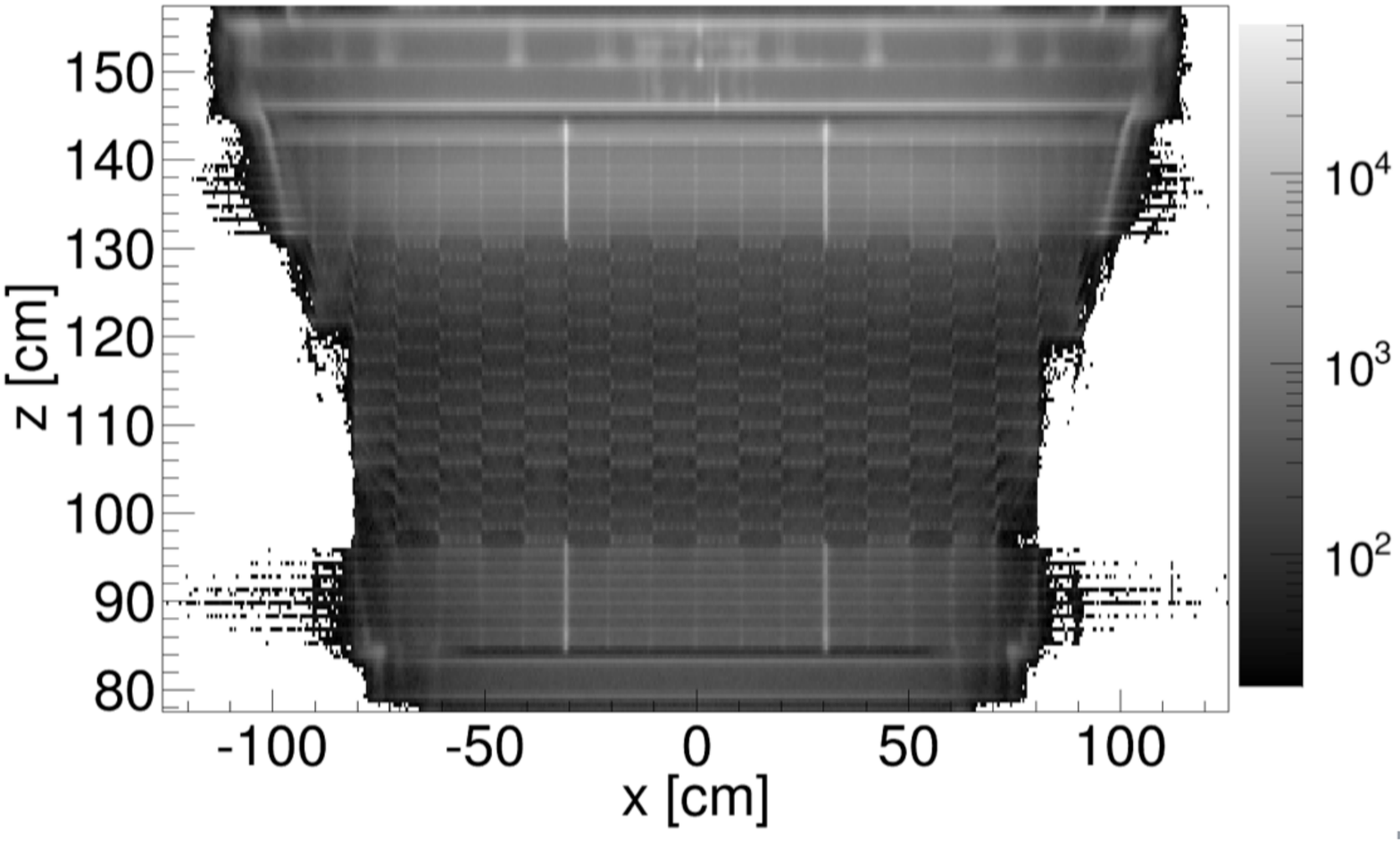}
\hspace*{-0.4cm}
\includegraphics[width=0.5\textwidth]{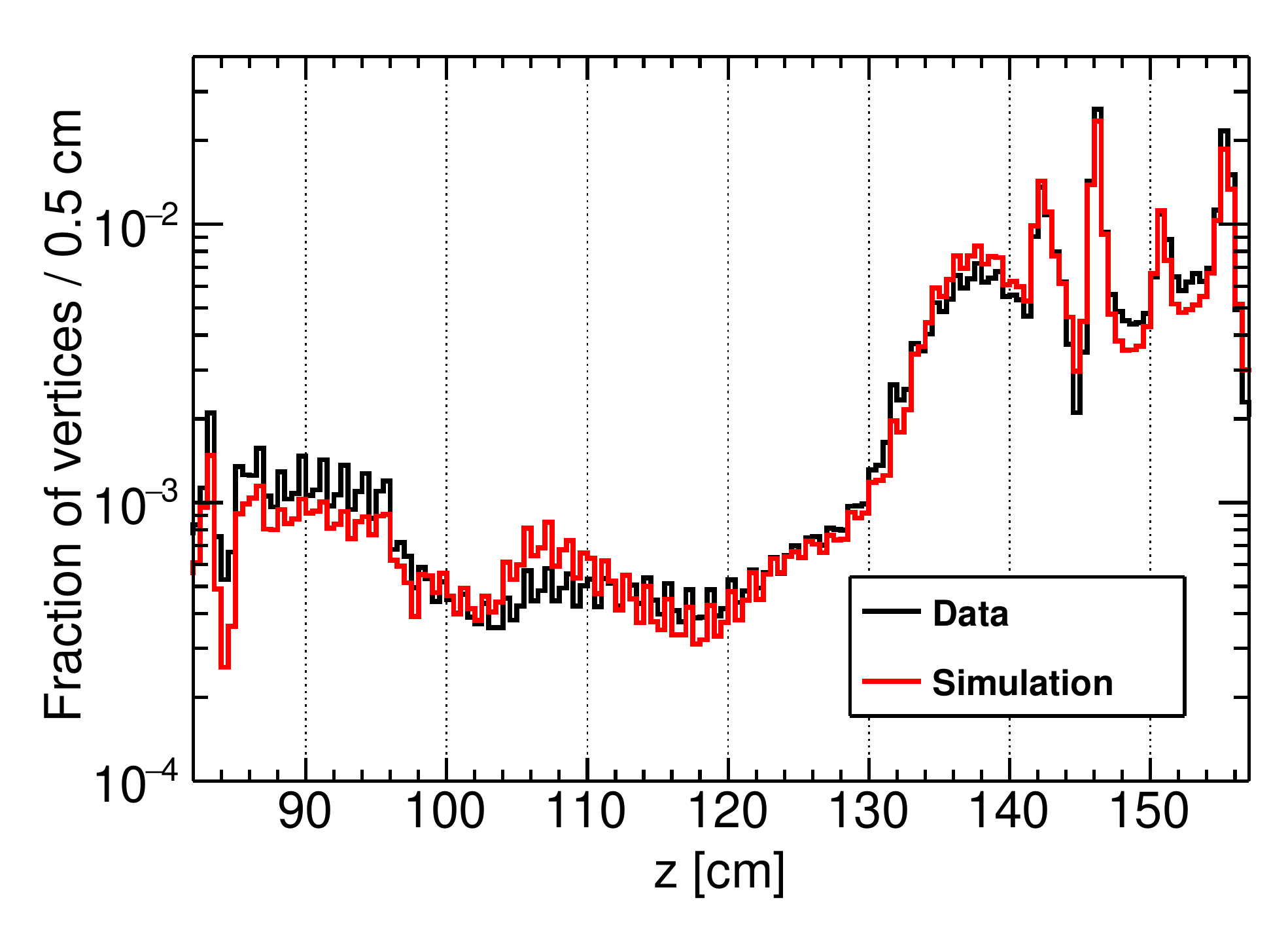}
\caption{Top: The number of reconstructed vertices ($\lvert y \rvert
  < 50$~cm) in AMS-02 data using TRD standalone tracks as visualized
  in XZ projection. Bottom: A comparison of the AMS-02 vertex
  probability distribution ($\lvert x,y \rvert < 20$~cm) in the TRD
  with simulation as a function of vertical position $z$.}
\label{fig:vertex}
\end{center}
\end{figure}

For a cross-check of the geometry description, we use rare events in
which hadronic interactions occur inside or close to the TRD,
resulting in secondary tracks that share a common vertex which can be
reconstructed by the standalone TRD tracking. The distribution of
interaction vertices reflects the material distribution in the
detector.  The amount of reconstructed vertices in AMS-02 data within
$\lvert y \rvert < 50$~cm in the TRD is shown in the top view of
Fig.~\ref{fig:vertex}. The TRD covers a vertical range of $z =
80$--$155$~cm, with the coordinate system defined as shown in
Fig.~\ref{fig:trd}. Even tiny details are visible demonstrating the
accuracy of the TRD standalone track reconstruction. In the bottom
view of Fig.~\ref{fig:vertex} the reconstructed vertex distributions
in the TRD are compared between AMS-02 and simulated data. The
fraction of vertices drops steeply in the lower part of TRD where the
probability to reconstruct sufficiently long TRD tracks decreases
strongly in the case of down-going primary particles. The lower the
$z$-position, the higher the probability that the vertex is based on
backsplash secondary tracks of a down-going or on secondary tracks of
an upward-going primary particle. These contributions can change the
relative fraction of vertices at different vertical positions but are
still reproduced well in the simulation.

In the TRD simulation, all energy depositions from primary and
secondary particles in a given straw are summed up. Then, the total
energy deposition amplitude is converted to an ADC value in a signal
digitization step, using a conversion factor of
$\approx\!47\,\mathrm{ADC}\,\mathrm{counts}/\mathrm{keV}$. This
procedure yields the ideal ADC value $\Delta E_\mathrm{ideal}$. The
straws act as ideal detectors, that is, detailed gas physics phenomena
like drifting of low-energy ionization electrons to the anode wire are
not simulated. The detailed simulation of gas physics would require
too much CPU time.

We find that the measured energy deposition is non-linear compared
with the ideal detector simulation. The non-linearity of the readout
electronics was measured in detail in the laboratory before the launch
of AMS-02, but its magnitude is not sufficient to explain the measured
difference to the ideal simulation. Therefore, gas effects have to be
approximated by an effective empirical correction applied in the
signal digitization procedure. We have found an empirical correction
that needs as few parameters as possible and is universal,
i.e. without explicit dependence on particle species.  We observe
several effects that we account for in the correction. All in all, the
effects are similar to space-charge effects
\cite{space-charge1969RScI...40.1216H, alice2004NIMPA.525..447A}.
First, we see a strong non-linearity for ions, which produce large
numbers of secondary electrons.  Second, the magnitude of
non-linearity is stronger the larger the relative fraction of primary
ionization.  This is likely due to the fact that drift electrons
originating from primary particle ionization arrive at the wire more
localized than that from secondary particle ionization. If a
significant fraction of the energy deposition comes from secondary
particles, which hit the tube at some distance from the primary track,
the drift electrons will be spread over a longer distance on the wire
resulting in a reduction of the non-linearity. To account for this
geometric effect, we will use the fraction of energy deposition that
originates from primary ionization, $f_\mathrm{prim} = \Delta
E_{\mathrm{ideal}}^{\mathrm{prim}} / \Delta E_{\mathrm{ideal}}$.
Third, the absorption of TR X-ray photons differs from continuous
ionization since a photon is absorbed in a single point from which all
electrons resulting from atom de-excitation drift together to the
wire. Therefore, we will use the fraction of transition radiation
$f_{\mathrm{TR}} = \Delta E_{\mathrm{ideal}}^{\mathrm{TR}} / \Delta
E_{\mathrm{ideal}}$ that contributed to the energy deposition in the
straw in the correction. The optimal corrections were found out by
trying several different approaches. To ensure that corrections are
smooth and behave as expected the simplest possible solutions were
selected.

\begin{figure}[!]
\begin{center}
\hspace*{-0.4cm}
\includegraphics[width=0.52\textwidth]{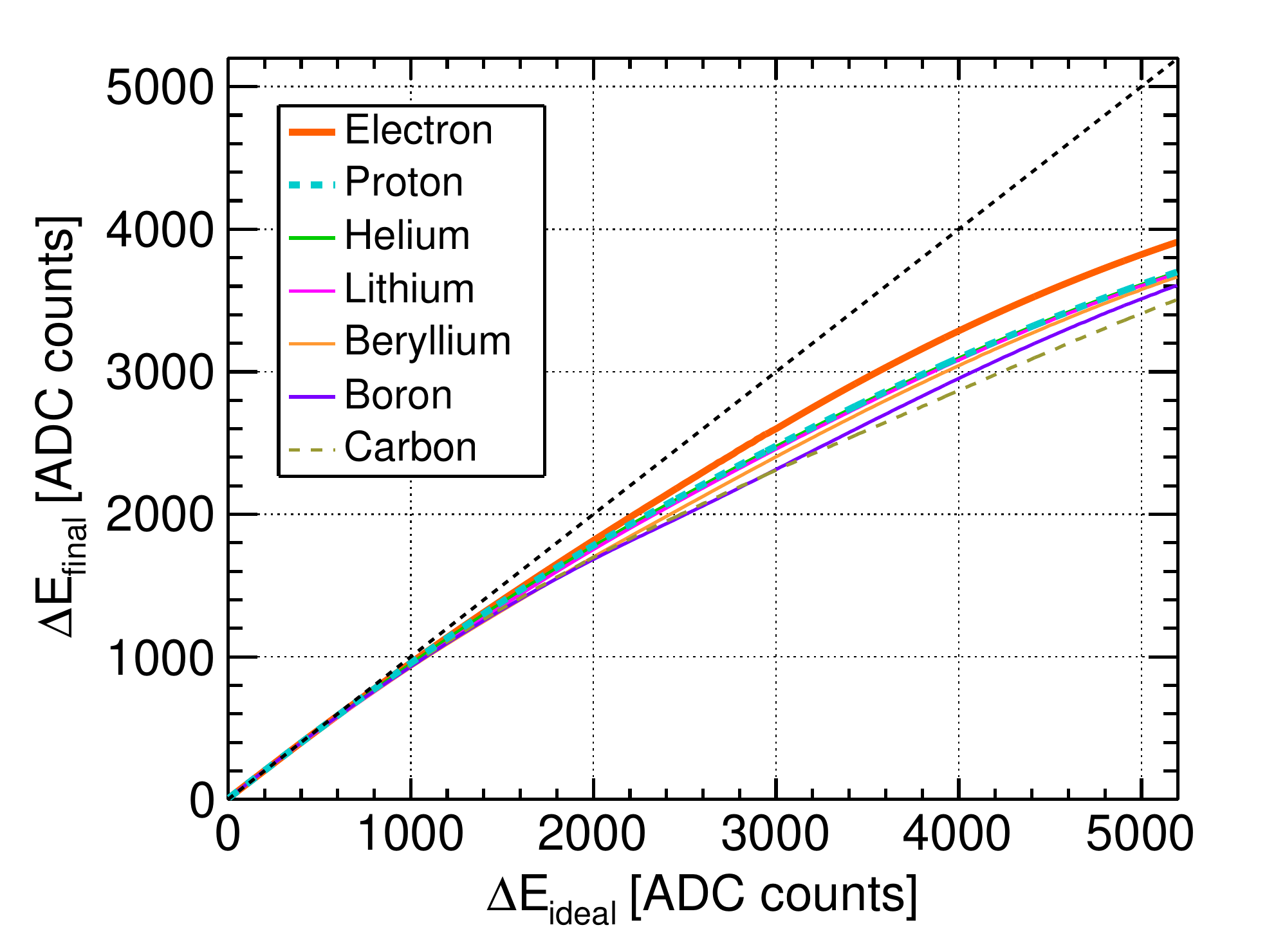}
\caption{The final mean non-linearity corrections applied to ideal
  simulated energy depositions $\Delta E_{\mathrm{ideal}}$ for each
  species in the rigidity range of $10$--$100$~GV.}
\label{fig:corrections}
\end{center}
\end{figure}

In the case of simulated data, all non-linearity effects present in
AMS-02 data and discussed above need to be included by applying
corrections to $\Delta E_\mathrm{ideal}$. In the first and strongest
correction, the ADC value is adjusted downwards according to the
second-order polynomial

\begin{equation}
\label{eq:prim_nonlinearity}
\Delta E_{\mathrm{corr}}'' =
  \Delta E_{\mathrm{ideal}} - 5.2 \times 10^{-5} \times \Delta E_{\mathrm{ideal}}^{2}, \quad \Delta E_{\mathrm{ideal}}< 2300,
\end{equation}

\noindent where $\Delta E$ is given in ADC counts. Beyond a value of
$2300$~ADC counts, the correction becomes linear and continues
tangentially. The amplitude is then further reduced depending on the
value of $f_\mathrm{prim}$:

\begin{equation}
\Delta E_{\mathrm{corr}}' = \Delta E_{\mathrm{corr}}'' - 0.85 \times \delta E \times f_{\mathrm{prim}}^2,
\end{equation}

\noindent where $\delta E = \Delta E_{\mathrm{ideal}} - \Delta
E_{\mathrm{corr}}''$ is the magnitude of the previous
correction. Here, we consider all energy depositions in gas except
absorption of TR photons as ionization. After the correction, some
compensation is done depending on the fraction of TR $f_{\mathrm{TR}}$,
defined as

\begin{equation}
\label{eq:tr_correction}
\Delta E_{\mathrm{corr}} =
\Delta E_{\mathrm{corr}}' + 0.0012 \times \delta E^2 \times f_{\mathrm{TR}},
\end{equation}

\noindent but not allowed to exceed $\delta E$. In the last step, a
correction for the non-linearity of the electronics, based on
laboratory measurements, is applied. It is approximated by the
third-order polynomial
\begin{equation}
\begin{split}
\Delta E_{\mathrm{final}} &= \Delta E_{\mathrm{corr}} + 1.39 \times 10^{-5} \times \Delta E_{\mathrm{corr}}^2 \\ &- 1.03 \times 10^{-8} \times \Delta E_{\mathrm{corr}}^3, \quad E_{\mathrm{corr}} < 5200,
\end{split}
\end{equation}

\noindent that continues linearly in the direction of the tangent
after $5200$~ADC counts. The electronics correction results in
$5$--$10$\% reduction in the end of ADC dynamic range. In the case of
pure ionization, the correction in Eq.~\ref{eq:prim_nonlinearity} has
the largest contribution. The final correction function defined above
is shown in Fig.~\ref{fig:corrections} for particle species with $Z
\le 6$. The correction function is independent of the particle species
but varies slightly and especially for $Z \ge 4$ because of different
primary ionization contributions. The final correction for electrons
differs clearly from that for other species because of the
TR-dependent compensation in Eq.~\ref{eq:tr_correction}. In general,
the TRD response is relatively linear up to $\sim1000$~ADC counts
after which non-linearity effects become important.

\section{Validation of TRD simulation with AMS-02 data}
\label{sec:comparison}

\begin{figure*}[!]
\begin{center}
\hspace*{-0.5cm}
\includegraphics[width=1.05\textwidth]{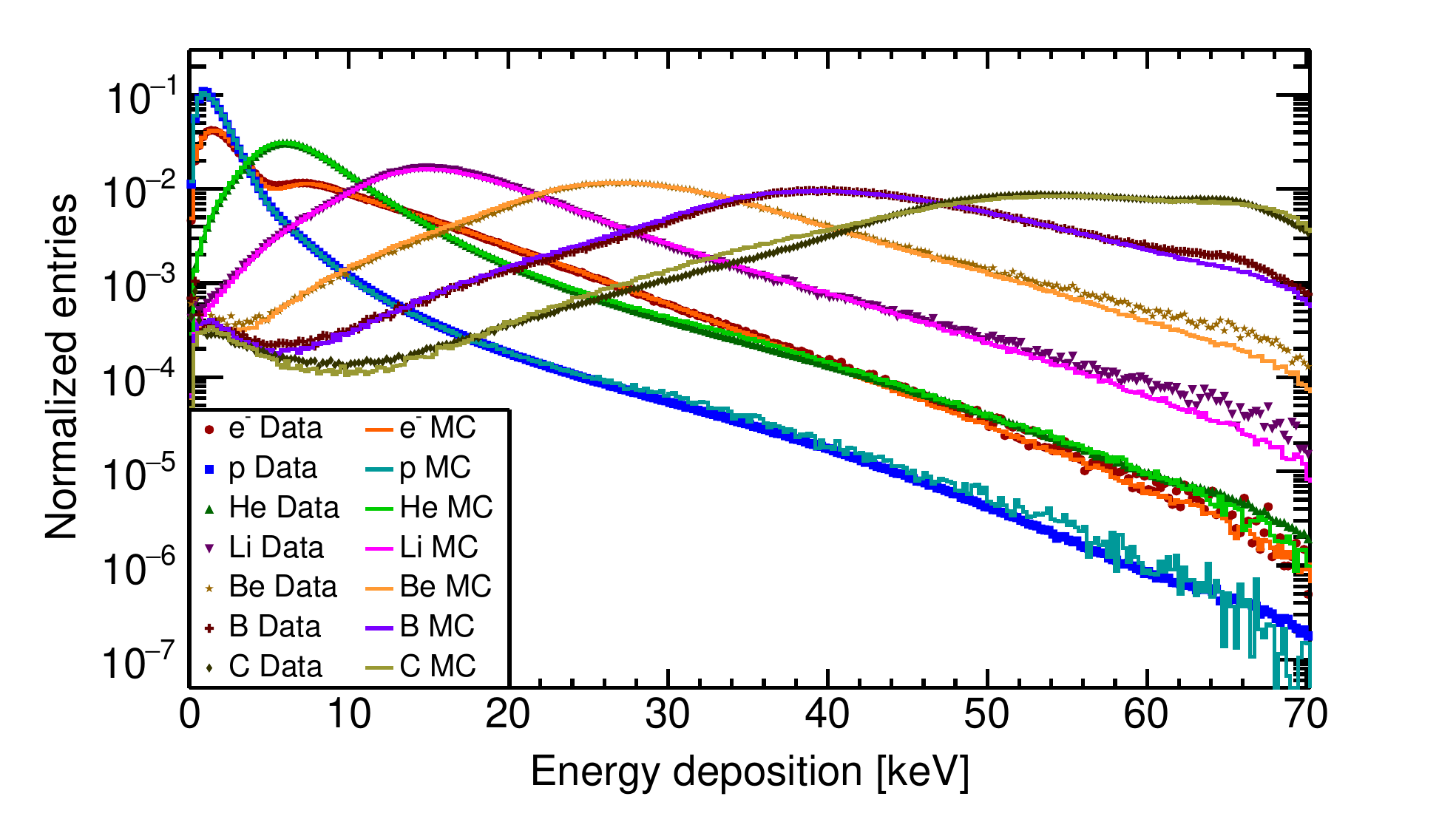}
\caption{The AMS-02 $\Delta E$ single-tube amplitude spectra of species $Z \le 6$
  compared with simulations in the rigidity range of $10$--$100$~GV.}
\label{fig:adcSpectraLog}
\end{center}
\end{figure*}

A precise description of particle energy deposition spectra is
important since they are used for particle identification and
background suppression in data analysis. The comparison between AMS-02
data and simulations for species $Z \le 6$ in the rigidity range of
$10$--$100$~GV is shown in Fig.~\ref{fig:adcSpectraLog}. Rigidity is
defined as momentum per charge. Clean samples of particle species in
the TRD were obtained by applying suitable selection cuts on
observables measured by the other subdetectors of AMS-02 in the
selected geometrical acceptance. In event selection, the xenon partial
pressure in the TRD was required to be within $\pm 30$~mbar of the
fixed $850$~mbar xenon pressure set in simulations to ensure similar
conditions for data comparison. The distribution in momentum $p$ of
the simulated primary particles was chosen to be proportional to
$p^{-1}$, and events were then re-weighted to match the actual
momentum spectra observed in cosmic rays.

\begin{figure}[!]
\begin{center}
\includegraphics[width=0.52\textwidth]{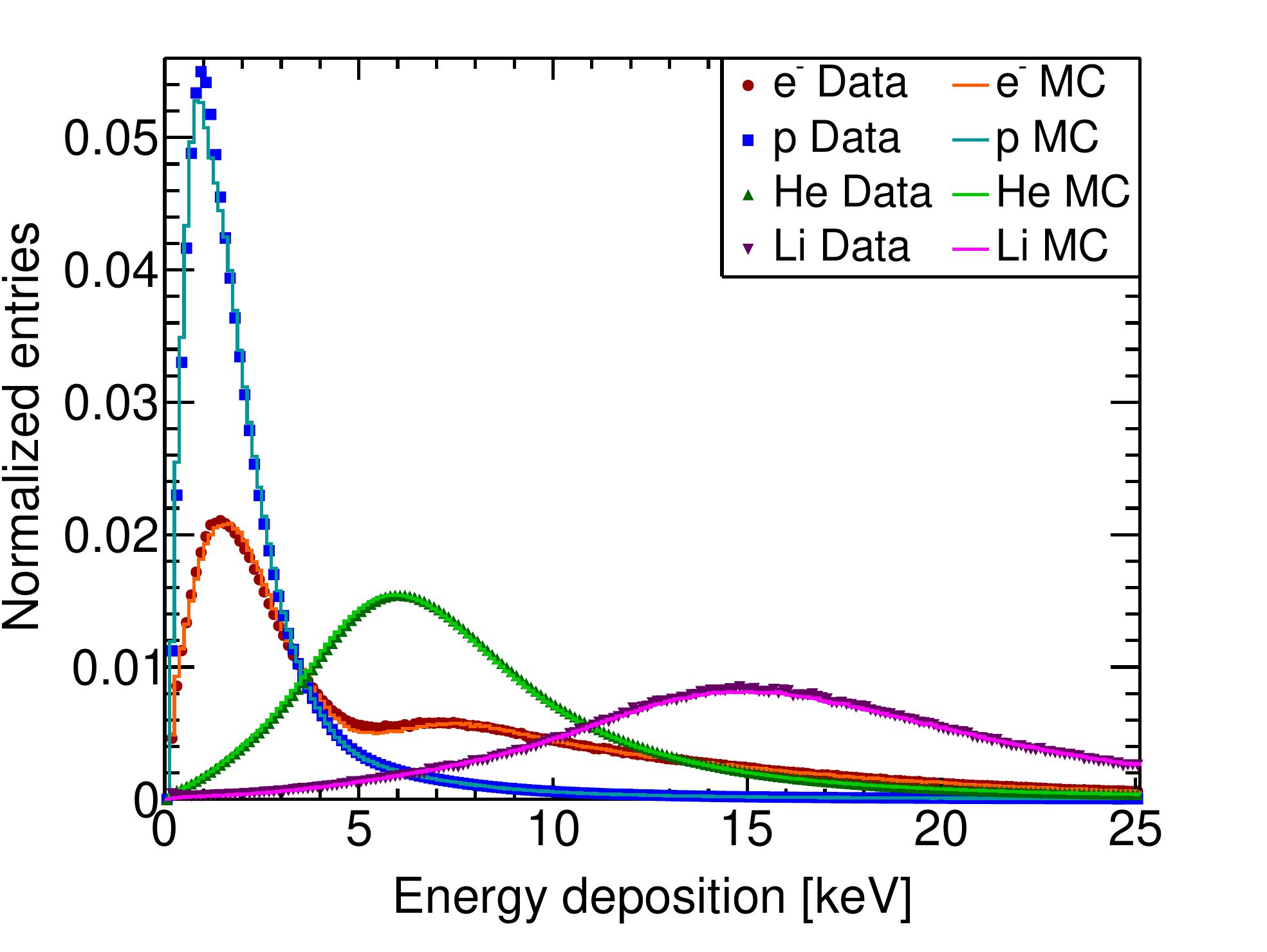}
\caption{The AMS-02 $\Delta E$ single-tube amplitude spectra of species $Z \le 3$
  compared with simulations in the rigidity range of $10$--$100$~GV in
  a linear scale. The distributions are normalized to unit integral.}
\label{fig:adcSpectraLin}
\end{center}
\end{figure}

The simulations show a very good agreement with experimental data over
a wide amplitude range for all the species. Small deviations from
AMS-02 data in the tails of the distributions can be partially
explained by the time-dependence of the straw pedestals, which changes
the effective end of the ADC range, as well as variations in gas
pressure which are not included in the simulation. As visible in
Fig.~\ref{fig:adcSpectraLin}, the shapes of the ionization peaks for
simulated protons and electrons differ only slightly from AMS-02 data.

\begin{figure}[!]
\begin{center}
\includegraphics[width=0.52\textwidth]{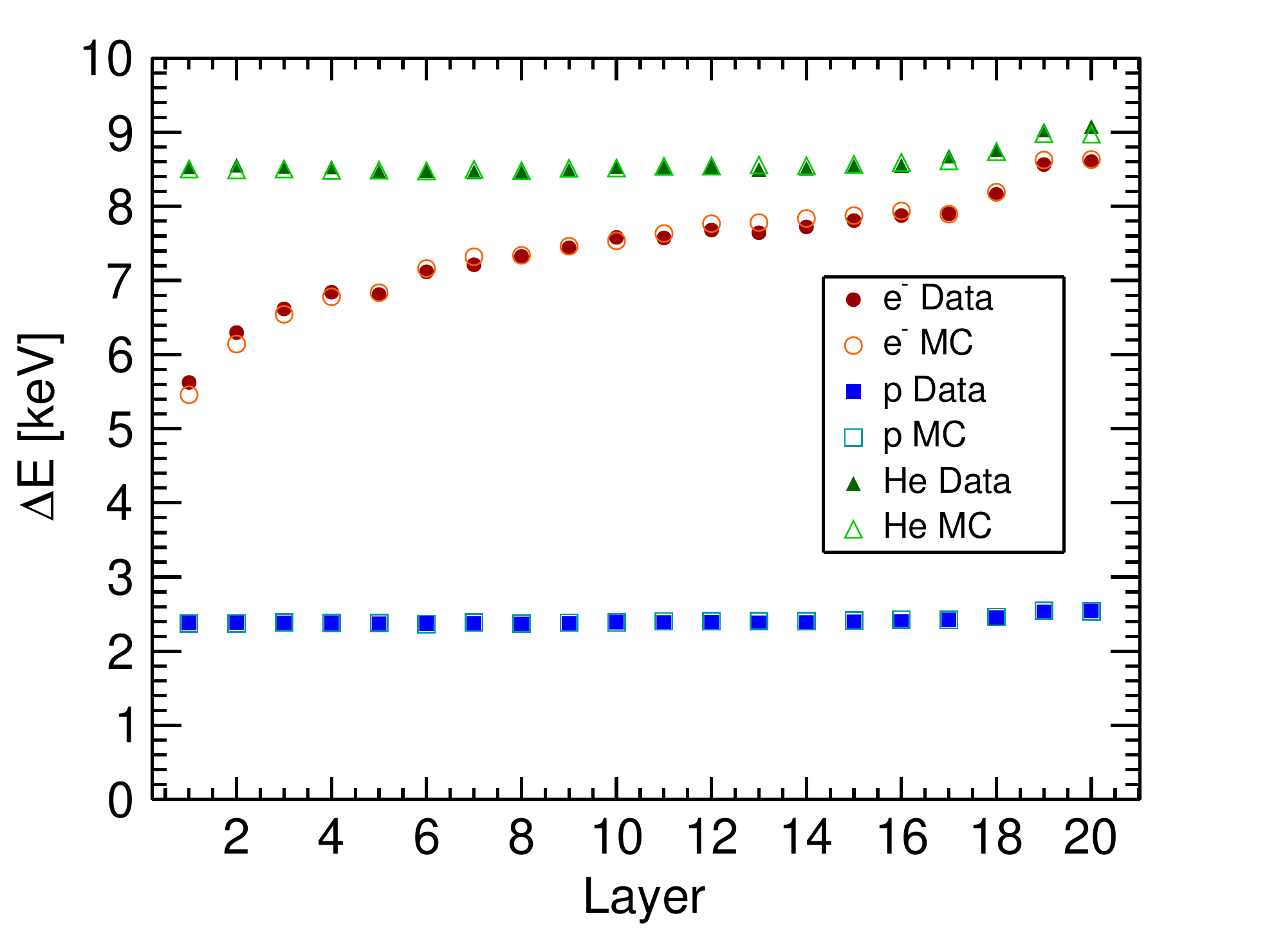}
\caption{The comparison of mean amplitudes in each TRD layer between
  AMS-02 and simulated electron, proton and helium nuclei data in the rigidity
  range of $10$--$100$~GV.
}
\label{fig:edepLayer}
\end{center}
\end{figure}

The accuracy of the TR simulation can be tested by comparing electron
mean energy depositions from layer to layer
(Fig.~\ref{fig:edepLayer}).  For protons and helium nuclei the mean
energy depositions in the TRD stay rather constant between the layers
but for electrons they increase with depth since the TR photons
accumulate as they can pass through several layers before being
absorbed. Since the orientation of modules changes above layers $5$
and $17$, there is less space for fleece radiator, which results in
slightly reduced energy deposition in those layers. In the two bottom
layers the mean energy deposition is enhanced because of backsplash
particles originating from interactions in the TRD lower honeycomb.

\begin{figure}[!]
\begin{center}
\includegraphics[width=0.52\textwidth]{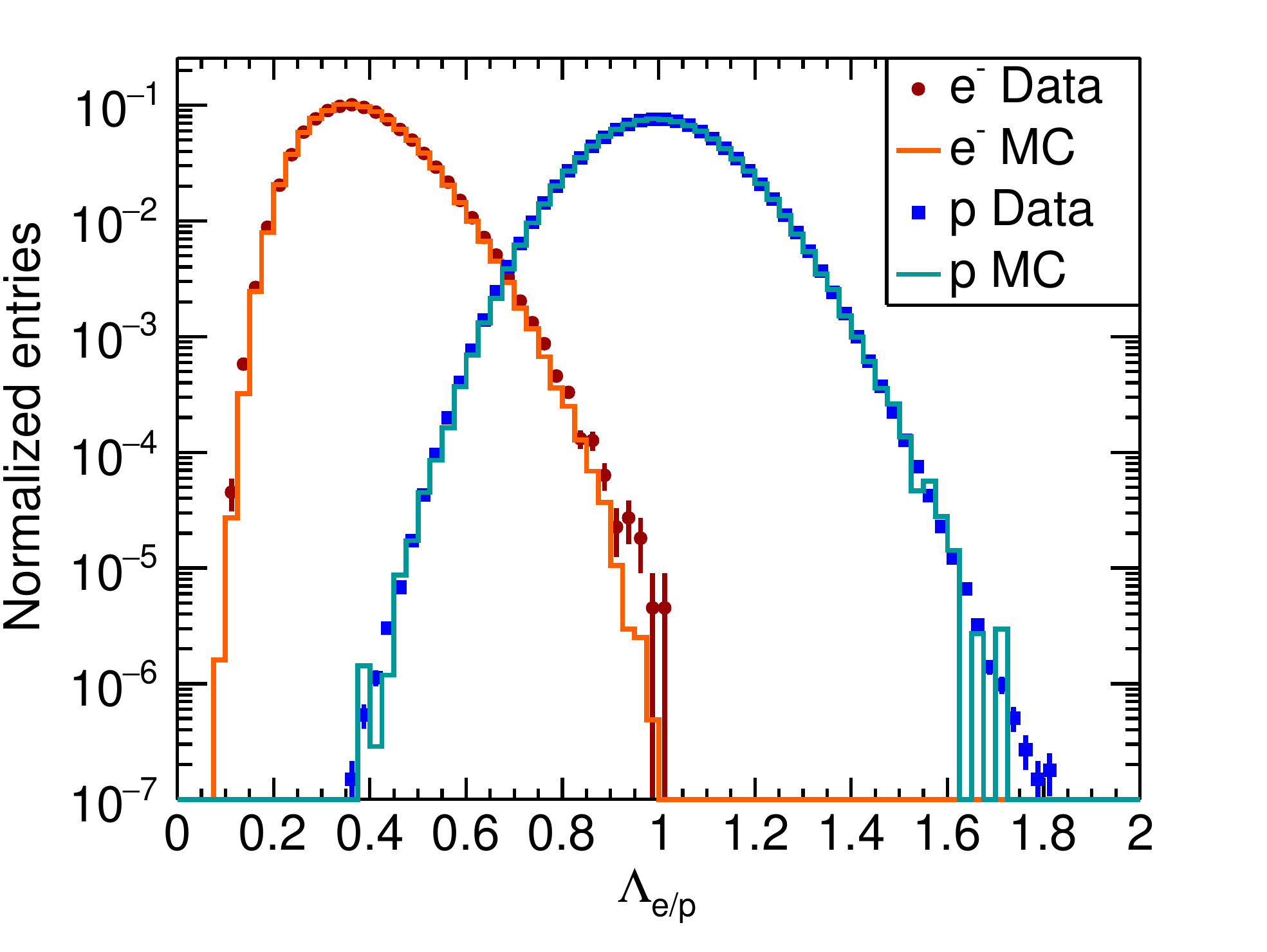}
\caption{The comparison between AMS-02 and simulated TRD
  $\Lambda_{\mathrm{e/p}}$ probability distributions of $10-100$~GV
  electrons and protons.}
\label{fig:ep-likelihoodRatio}
\end{center}
\end{figure}

\begin{figure}[!]
\begin{center}
\includegraphics[width=0.52\textwidth]{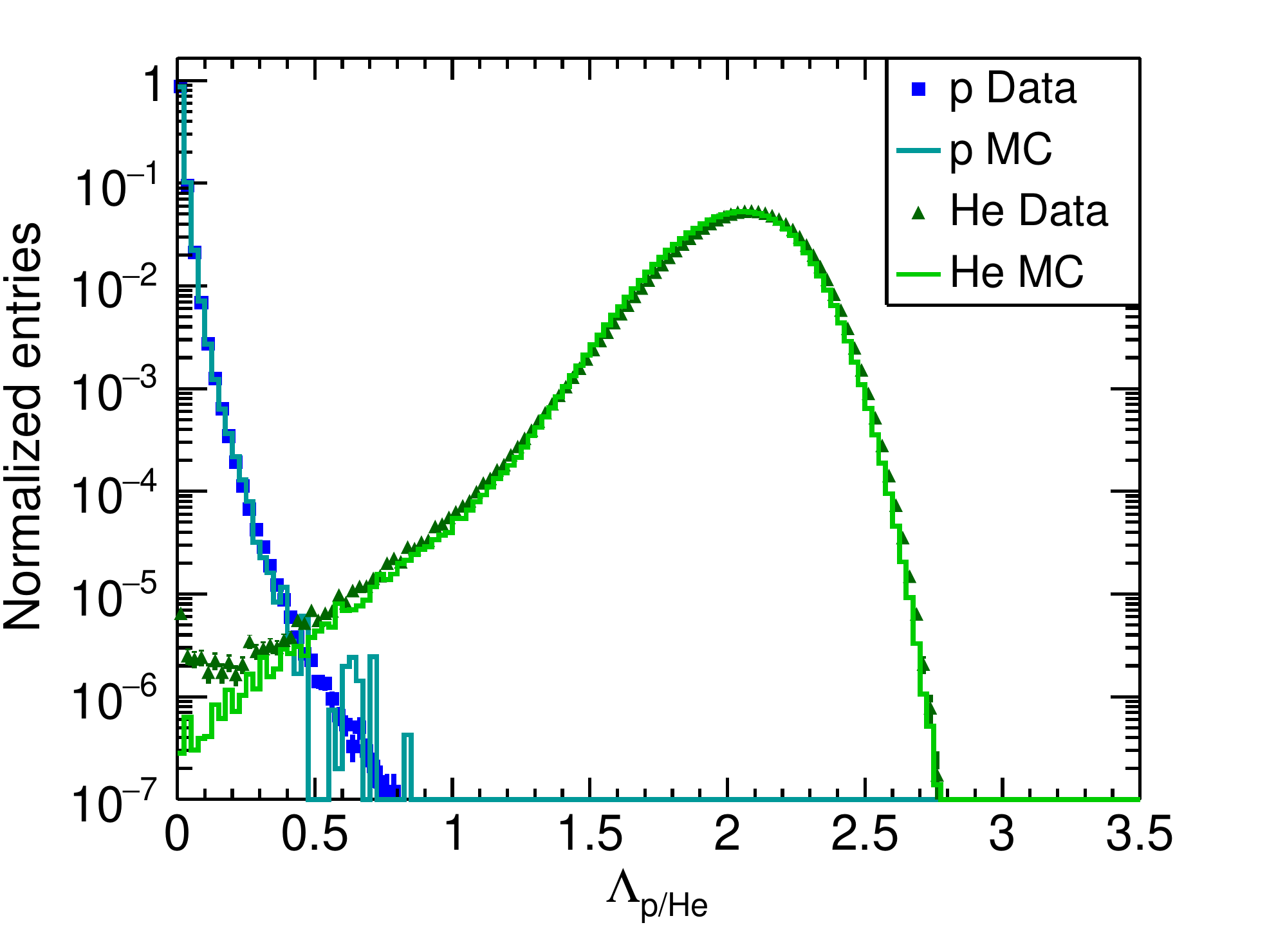}
\caption{The comparison between AMS-02 and simulated TRD
  $\Lambda_{\mathrm{p/He}}$ probability distributions of $10-100$~GV
  protons and helium nuclei.}
\label{fig:pHe-likelihoodRatio}
\end{center}
\end{figure}

For particle identification purposes, one employs a TRD estimator
$\Lambda_{\mathrm{i/j}}$ calculated as a likelihood
ratio between two particle hypotheses $i$ and $j$,

\begin{equation}
  \Lambda_{\mathrm{i/j}} := -\log \left( \frac{\mathcal{L}_{i}}{\mathcal{L}_{i} + \mathcal{L}_{j}} \right),
\end{equation}

\noindent in which the likelihood $\mathcal{L}_{k}$ for each particle
hypothesis $k$ is calculated from the measurements in
$N=\mathcal{O}(20)$ layers along one particle track in the TRD according to

\begin{equation}
\mathcal{L}_k = \sqrt[N]{\prod_{l = 1}^{N} p_{k} (R,\ell,P_\mathrm{Xe},dE,dx)},
\end{equation}

\noindent where $p_{k}$ is the probability for a particle of species
$k$ and rigidity $R$ to give an energy deposition $dE$ in layer $\ell$
at a xenon partial pressure of $P_\mathrm{Xe}$ over the measured
pathlength $dx$. Lookups for the probability functions $p_k$ are
created from AMS-02 data in a dedicated calibration analysis. The
comparison of the $\Lambda_{\mathrm{e/p}}$ distributions for protons
and electrons and of the $\Lambda_{\mathrm{p/He}}$ distributions for
protons and helium between simulations and AMS-02 data is shown in
Figs.~\ref{fig:ep-likelihoodRatio} and
\ref{fig:pHe-likelihoodRatio}. The simulation reproduces AMS-02 data
with high accuracy.

\begin{figure}[!]
\begin{center}
\hspace*{-0.2cm}
\includegraphics[width=0.53\textwidth]{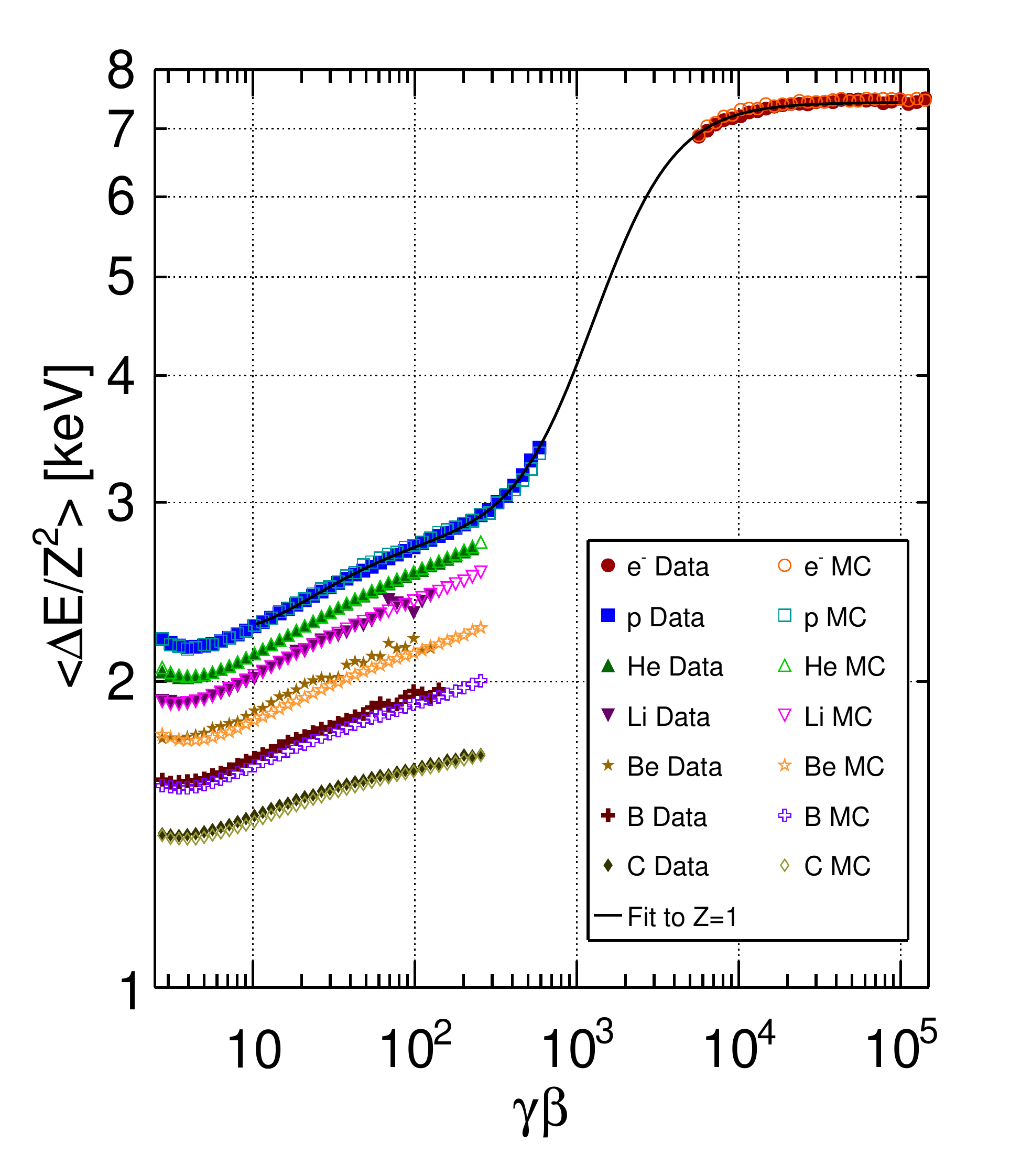}
\caption{The average mean $\Delta E$ per $Z^2$ of all hits in $20$
  layers shown as a function of $\gamma$. The solid line
  represents a parametrized fit to $Z=1$ AMS-02 datasets.}
\label{fig:meanResponse}
\end{center}
\end{figure}

Another useful analysis variable is the TRD mean response, which is
calculated as the mean $\Delta E$ of all hits that were associated
with the primary track. The average mean response per $Z^2$ as a
function of $\gamma\beta$ for particle species $Z \le 6$ is shown in
Fig.~\ref{fig:meanResponse}, where $\gamma\beta$ is calculated for
electrons from ECAL energy and for the rest of the particle species
from the reconstructed rigidity. The mean response per $Z^2$ gets
systematically reduced with increasing charge because of the
non-linearity effects, which get stronger the higher the energy
deposition in a straw. The onset of TR at $\gamma\beta\sim 500$ is
clearly visible. A modified Fermi function was fitted to the $Z=1$
AMS-02 datasets to demonstrate that they can be described universally.

\section{Conclusions}
\label{sec:conclusions}

We have shown that the geometry of the AMS-02 TRD is accurately
modeled in our simulation. After optimization of the TRD simulation
parameters and implementation of empirical non-linearity corrections,
the TRD amplitude and likelihood distributions of all relevant
particle species up to carbon nuclei are described at remarkable
precision over a wide rigidity range. The non-linearity corrections,
which are applied to simulated data in the digitization phase, play an
important role for achieving a realistic detector simulation. The
correction functions are independent of $Z$ and they are motivated by
physics principles. All in all, the TRD simulation works very well and
provides important input to data analyses.

\section*{Acknowledgments}
We are grateful to our colleagues of the AMS Collaboration, in
particular those involved in the design, construction and operation of
the TRD. Simulations were partially performed with computing resources
granted by JARA-HPC from RWTH Aachen University under project
jara0052.





\biboptions{sort&compress}

\end{document}